\newcommand{\simge}{\mbox{$\stackrel{>}{_{\sim}}$}}
\documentstyle[epsfig,aaspp4]{article}
\lefthead{Tuthill et al.}
\righthead{Sub-arcsecond structure of R~Aqr}

\begin{document}
%_____________________________________TITLE PAGE___________________________
\title{Near and mid-IR sub-arcsecond structure of 
           the dusty symbiotic star R~Aqr}
\author{P.G. Tuthill\altaffilmark{1,2},  W.C. Danchi\altaffilmark{1},  
       D.S. Hale\altaffilmark{1}, J.D. Monnier\altaffilmark{1,3}, 
       \and C.H. Townes\altaffilmark{1} 
} 
\altaffiltext{1}{Space Sciences Laboratory, University of California, Berkeley,
                 Berkeley,  CA  94720-7450, USA}
\altaffiltext{2}{Chatterton Astronomy Department, School of Physics, University of
                Sydney, NSW 2006, Australia }
\altaffiltext{3}{Smithsonian Astrophysical Observatory, 60 Garden Street, 
                Cambridge, MA, 02138, USA}
%\email{gekko@physics.usyd.edu.au, wcd@ssl.berkeley.edu, david@isi9.mtwilson.edu,
%       jmonnier@cfa.harvard.edu, cht@ssl.berkeley.edu}
%
%
%____________________________________ABSTRACT PAGE_________________________
\begin{abstract}

The results of a high-resolution interferometric 
campaign targeting the symbiotic long-period variable (LPV) R~Aqr are reported.
With both near-infrared measurements on baselines out to 10\,m and
mid-infrared data extending to 32\,m, we have been able to measure
the characteristic sizes of regions from the photosphere of the
LPV and its extended molecular atmosphere, out to the cooler
circumstellar dust shell.
The near-infrared data were taken using aperture masking interferometry
on the Keck-I telescope and show R~Aqr to be partially resolved
for wavelengths out to 2.2\,$\micron$ but with a marked enlargement, 
possibly due to molecular opacity, at 3.1\,$\micron$.
Mid-infrared interferometric measurements were obtained with the
U.C. Berkeley Infrared Spatial Interferometer (ISI) operating at 
11.15~$\micron$ from 1992 to 1999.
Although this dataset is somewhat heterogeneous with incomplete
coverage of the Fourier plane and sampling of the pulsation cycle, clear 
changes in the mid-infrared brightness distribution were observed, both as a 
function of position angle on the sky and as a function of pulsation phase.
Spherically symmetric radiative transfer calculations of uniform-outflow
dust shell models produce brightness distributions and spectra which
partially explain the data, however limitations to this approximation are 
noted.
Evidence for significant deviation from circular symmetry was found in the 
mid-infrared and more tentatively at 3.08\,$\micron$ in the near-infrared, 
however no clear detection of binarity or of non-LPV elements in the 
symbiotic system is reported.

\end{abstract}

%_______________________________________INTRODUCTION_______________________
\section{Introduction}

R~Aqr is a mass-losing long-period variable (LPV) star believed to be in a symbiotic 
system with an obscured hot companion whose presence is betrayed by nebular 
emission lines.
Long noted for its peculiar elliptical visible nebulosity extending up to 2' 
(first reported in Lampland 1923), this star has recently been subject to
intense scrutiny with the latest generation of astronomical instrumentation
both on the ground and in space.
Much of this attention can be traced to the detection in the optical
(\cite{WalGrens80}; \cite{Herbig80}) and radio (\cite{Sopka_et82})
of what is believed to be the nearest astrophysical jet to the earth
($\sim 200$\,pc; \cite{VanL_et97}; \cite{HPL97}). 
Thought to originate in the accretion disk around an unseen hot sub-dwarf,
this jet has been studied extensively at UV, optical and radio
wavelengths (e.g. \cite{BVP92}; \cite{PH94}; \cite{LJ92}).

A difference in location of the peak intensity of the $\nu =1, J = 1-0$ SiO 
maser line and the nearby lineless continuum from 7\,mm VLA maps was
reported by Hollis, Pedelty and Lyon (1997).
These authors interpreted the SiO peak as the location of the LPV and the 
continuum peak marking the accretion disk of the companion, and were thus
able to derive a binary separation of $55\pm2$\,mas with a position angle 
of $18^\circ \pm 2^\circ$.

With its ability to penetrate layers of obscuring dust, the infrared might 
be thought an ideal wavelength to image the inner regions of this system, 
from photospheres of the stars to dust and jets in their immediate surroundings.
However the high angular resolutions required and the presence of the
enormously luminous LPV have limited the efficacy of this approach.
Infrared interferometric measurements with the IOTA array (\cite{vanB96}) 
yielded photospheric diameter measurements of $\sim$14--15\,mas for the LPV  
but no evidence for an additional component.
Speckle observations at J band (\cite{KMC94}) suggestive of an elongated
image were interpreted in terms of a distorted inner dust shell, possibly 
due to the presence of the companion.
Anandarao \& Pottasch (1986) were able to match the near- to far-IR
fluxes using a two-temperature dust shell model, while more recently
Le Sidaner \& Le Bertre (1996) were able to fit similar spectral energy 
distributions with a simple uniform outflow model. 

In the following section of this paper, we briefly describe our 
observational methods and apparatus for obtaining near- and mid-IR
interferometric measurements. 
In Section~3 the experimental results are given, together with 
discussion of model fitting and physical interpretation. 
A brief summary of our conclusions is found in Section~4.

%_______________________________OBSERVATIONS____________________________
\section{Observations}

A brief synopsis of the observing techniques used to secure the near-
and mid-IR interferometric measurements is given below.
Near-IR data was taken at the Keck-I telescope in 1998 July and 1999 January 
covering 4 narrowband wavelengths between 1.25 and 3.08~$\micron$.
Visibility data at 11.15~$\micron$ were taken with the Infrared Spatial
Interferometer (ISI) over the period 1992 to 1999 with various baselines
sampling the visibility function.

\subsection{Near-Infrared Interferometry} 

Near-IR observations at the Keck-I telescope have utilized the technique
of aperture masking interferometry, in which a mask is placed over the
telescope pupil so as to only pass light from selected regions, in effect
transforming the telescope into an array of small subapertures.
Starlight passing through the mask, in this case a 21-hole non-redundant
configuration, was brought to a focus in the Near InfraRed Camera 
(\cite{ms94}; \cite{matthews96}), a $256 \times 256$ InSb array with a 
pixel scale of 20.57\,milli-arcsec/pixel.
At focus, an interference pattern is formed containing sets of fringes
at various spatial frequencies corresponding to baselines in the pupil.
Subsequent analysis of datasets consisting of 100 rapid-exposure 
($\sim 140$\,msec) frames using Fourier techniques enables recovery of
the visibility amplitudes and closure phases corresponding to the complex 
visibility function of the object.
With the exception of the sparse telescope pupil which has been shown
to confer signal-to-noise advantage for bright targets, the observational
and data processing techniques used were very similar to those utilized
in speckle interferometry (for a review, see \cite{Roddier88}).

After calibration utilizing nearly-contemporaneous observations of nearby
point-source stars, the visibility data could be interpreted in a
number of ways, including the fitting of model brightness distributions.
The recovery of diffraction-limited maps was also possible with the
help of self-calibration methods such as the Maximum-Entropy Method
(\cite{gs84}; \cite{sivia87}) or CLEAN algorithm (\cite{hogbom74}).
The signal-to-noise advantages of sparse-aperture observations for bright
sources have been exploited previously (e.g., \cite{baldwin86}; 
\cite{Haniff_et87}; \cite{Roddier88}), while a more detailed description of
this particular apparatus and technique may be found in Monnier et al. (1999) 
and Tuthill et al. (1999a).
An observing log of near-IR observations is given in Table~1 showing the
dates, filters and stellar phases (\cite{aavso99}) pertaining to our two 
sets of measurements in 1998 June and 1999 January.

%%%%%%%%%%%%%%%%%%%%%%%%%%%%
% Table One - Nirc Journal %
%%%%%%%%%%%%%%%%%%%%%%%%%%%%
\begin{deluxetable}{cccc}
\tablewidth{0pt}
\tablecaption{Journal of Near-IR Keck Observations}
\tablehead{
 \colhead{Date} & \colhead{Filter Wavelength} & 
 \colhead{Filter Bandwidth} & \colhead{Stellar Phase} \nl
 &  \colhead{($\mu$m)} & \colhead{($\mu$m)}  & \nl
}
\startdata
1998 Jul 05 & 1.236 & 0.011 & 0.12 \nl
1998 Jul 05 & 1.647 & 0.018 & 0.12 \nl
1998 Jul 05 & 2.260 & 0.053 & 0.12 \nl
1998 Jul 05 & 3.082 & 0.101 & 0.12 \nl
1999 Jan 04 & 2.260 & 0.053 & 0.68 \nl
1999 Jan 04 & 3.082 & 0.101 & 0.68 \nl
\enddata
\end{deluxetable}

\subsection{Mid-Infrared Interferometry} 

Our mid-IR visibility data were obtained at 11.15\,$\micron$ with the
U.C. Berkeley ISI, a two-element heterodyne stellar interferometer located
on Mt. Wilson, CA.
Both telescopes are mounted within movable semi-trailers which allowed periodic 
reconfiguration of the baseline from 4 to 32\,m over the course of these measurements.
Detailed descriptions of the apparatus and recent upgrades can be found
in Bester et al. (1990; 1994), Lipman (1998) and Hale et al. (1999).
A journal of R~Aqr observations taken with the ISI is provided in Table~2, 
which shows observations broken into seven separate observing epochs
labeled for convenience `A' through `G'.
As individual visibility measurements taken with the ISI can have low 
signal-to-noise, we have averaged together measurements taken over these 
periods with the result that the observing parameters (such as sky position 
angle or stellar variability phase) pertinent to any single datum can be 
traced back to the range of values given in Table~2.
Observations of K giant stars $\alpha$~Tau and $\alpha$~Boo were utilized
to monitor system visibility drifts which could be as large as 
15\% from year to year, thus ensuring reliable calibration (to better
than 5\%) of science targets as is discussed at some length in 
Danchi et al. (1990; 1994)

%%%%%%%%%%%%%%%%%%%%%%%%%%%
% Table Two - ISI Journal %
%%%%%%%%%%%%%%%%%%%%%%%%%%%
\begin{deluxetable}{cccccc}
\tablewidth{0pt}
\tablecaption{Journal of Mid-IR ISI Observations}
\tablehead{
\colhead{Run Label} & \colhead{Date} & \colhead{Tel Separation} & 
 \colhead{Spatial Frequency} & \colhead{Position Angle\tablenotemark{a}} &
 \colhead{Stellar Phase} \nl
 &  & \colhead{(meters)} & 
   \colhead{($10^{5}$ $\rm{rad}^{-1}$)} &
   \colhead{(Degrees)} &   \nl
}
\startdata
A & Sep - Oct 92 & 13.09 &  7.00 - 11.75 & 105 - 112 & 0.68 - 0.74 \nl
B \tablenotemark{b} & Jun - Jul 94 & 32.16 & 16.41 - 28.54 & 111 - 114 & 0.36 - 0.39 \nl
C & Sep - Oct 94 &  4.02 &  1.96 -  3.49 &  67 -  87 & 0.61 - 0.68 \nl
D & Aug - Sep 97 &  4.02 &  1.94 -  3.56 &  67 -  88 & 0.37 - 0.40 \nl
E & Oct - Nov 97 & 16.07 &  8.37 - 14.39 & 130 - 125 & 0.50 - 0.58 \nl
F & Jul - Aug 98 &  9.60 &  2.57 -  8.51 & 145 - 118 & 0.24 - 0.29 \nl
G & May - Jun 99 &  4.02 &  2.14 -  3.47 &  71 -  87 & 0.02 - 0.07 \nl
\enddata
\tablenotetext{a}{Position angle is measured in degrees East of North}
\tablenotetext{b}{One visibility datum at $28.54\times 10^{5}$ $\rm{rad}^{-1}$
 included in this set was taken in Nov~93 at a stellar phase of 0.76}
\end{deluxetable}

%___________________________________RESULTS_________________
\section{Results and Discussion}

\subsection{Near-Infrared Visibility Measurements} 

The principal aim of the near-IR program was to attempt to detect the 
presence of hot circumstellar material, either in the immediate environment
of the LPV or associated with the symbiotic companion.
The full two-dimensional Fourier coverage out to 10\,m baselines afforded by
the masking experiment has proved highly successful at recovering asymmetric
circumstellar structures in other systems (e.g., \cite{Monnier_et99}; 
\cite{TMD104_99}).
However, in the case of R~Aqr, the optical depth to the star is relatively
low in the near-IR ($\tau =$ 0.01 at 2.2\,$\mu$m), so that the star greatly 
outshines the contribution from the dust shell.
No circumstellar features were detected at any wavelength in the near-IR from J 
through K bands, and we are able to place an upper limit of around 
$\Delta M $\simge~5~magnitudes for the relative brightness in the near-IR 
of any such companion.

Failure to detect a second star may seem surprising in view of previous
indications of a hot companion.  Resolution of the interferometry 
is about 15 mas (FWHM) and the projected distance to a second
star might possibly be less than this.  However, Hollis et al. (1997)
reported indications in 1996 of a companion separated by 55 mas,
or 11 AU assuming the distance to R Aqr is 200 pc.  The orbital period
was estimated to be 44 years, so that its distance during the 
present measurements would have to be comparable with 55 mas and
easily resolved.  However, a near-IR intensity less than 1\% of that of
the R Aqr Mira is in fact not unreasonable.  The LPV has a diameter
of 15 mas, or 3 AU, whereas a hot star even as large as our Sun would
have a diameter at least 100 times smaller, and even allowing for its higher
irradiance, a factor of $\sim 10^3$ less flux should result in the near-IR.

Although the stellar disk of R~Aqr itself was not well resolved 
with the baselines available, its size was determined by fitting
the available data.
Figure~1 displays visibility curves for four different colors and
two observing epochs (see Table~1).
In order to increase the signal-to-noise in the plotted data, the two-dimensional
visibility function for each wavelength has been azimuthally averaged, allowing 
visibilities to be plotted as a function of baseline length. 
On this occasion, additional correction had to be made for the size of the
calibrator star 30~Psc, which was itself a late-type (M3III) variable expected 
from effective temperature and photometric arguments to present
an angular diameter of 7.52\,mas (\cite{OH82}).
Also overplotted on Figure~1 are the best-fitting circular uniform disk model
profiles with diameters given in the figure key.
As short baseline visibilities can suffer from poor calibration due to seeing
changes (the well-known `seeing spike' problem {\sl c.f.} Tuthill et al. 1999a),
we have chosen to fit data only at spatial frequencies higher than 
$2 \times 10^5$\,rad$^{-1}$.

%\medskip
%\vbox{
\begin{figure}
\begin{center}
\epsfig{file=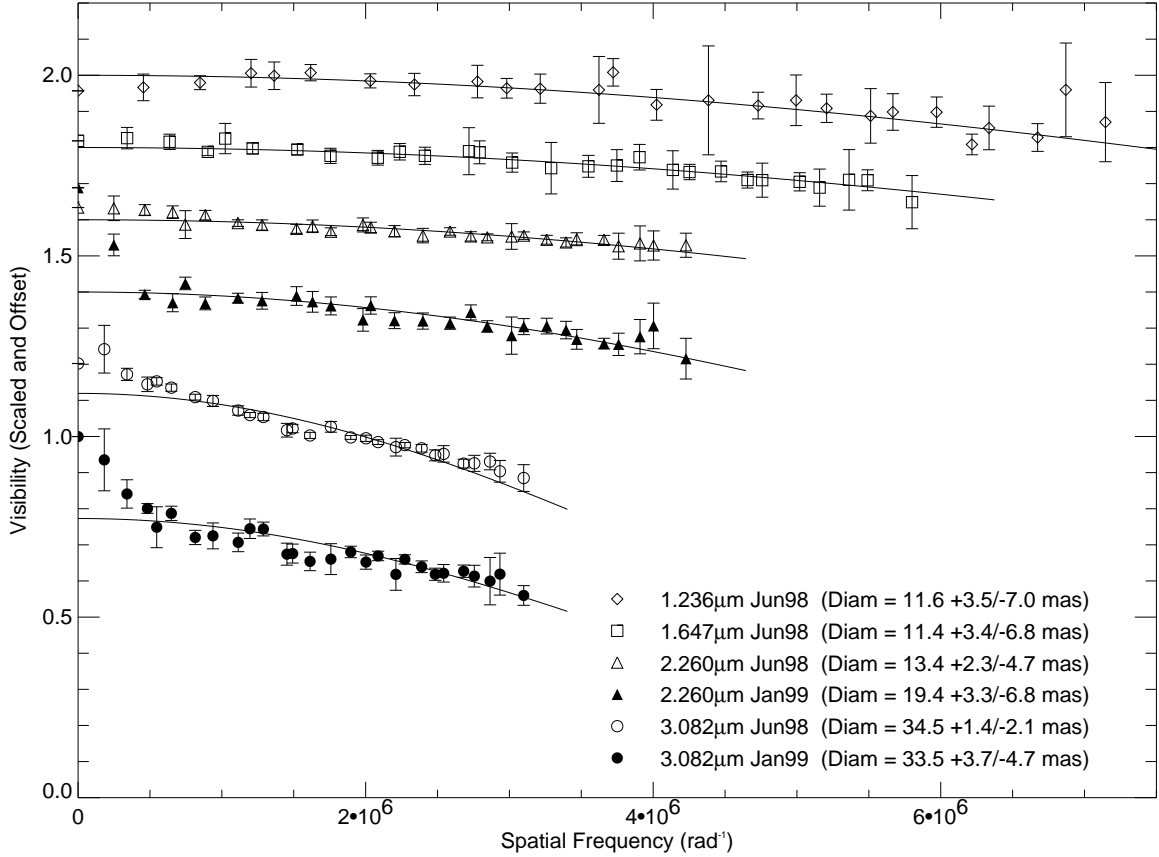,width=12cm,angle=90}
\smallskip
\caption{
%\noindent
%\footnotesize Fig 1.--- 
Azimuthally averaged visibility data for R~Aqr, plotted as a function of spatial 
frequency, obtained from near-IR observations at the Keck telescope.
Open symbols denote data taken on 1998~July~05 (4 colors), while filled symbols 
show data taken on 1999~January~04 (2 colors).
Visibilities have been offset by multiples of 0.2 in order to separate the various 
datasets on the plot.
Overplotted solid lines are best-fitting circular uniform disk model profiles with 
diameters given in the figure key.
These models and the asymmetric errors are further discussed in the text.
\label{fig:nearir}}
\end{center}
\end{figure}

It should be noted that over the wavelength range 1.236 -- 2.260\,$\mu$m (upper
four sets) the stellar diameter is really at the limit of our detection. 
Visibility curves have only dropped some $ \sim 10 -- 20$\% as compared to an 
unresolved source.
Although the formal diameter errors to the plotted data are small, the larger
errors given in Figure~1 reflect systematic sources of uncertainty due to
seeing-related mismatches between the source and point calibrator measurements.
The likely spread of this dominant term in the uncertainty was determined by 
examination of large volumes of additional calibration data taken with the same 
observing parameters.
Asymmetries in the errors are a consequence of the nonlinear relationship between
visibilities and angular diameters -- symmetric error bars on a visibility datum
will result in asymmetric diameter errors for such barely-resolved targets.
Despite these large uncertainties, we find good agreement with the published K 
band angular diameters of van Belle et al. (1996) of 14.95 and 14.06\,mas in 
1995~Jul~11 and 1995~Oct~07.
We note also, in passing, that our 1999 January 2.26\,$\mu$m diameter taken after minimum 
light is larger than the 1998 June observation near to maximum.
Such cycle-dependent size behavior has been studied by Burns et al. (1997) for the case 
of R~Leo, and although our enlargement is in rough accord with this earlier work, our
large errors and poor phase sampling cast doubt on the tentative detection of such 
dynamical changes here.

The masking experiment at the Keck telescope recorded full two-dimensional Fourier
information out to the maximum possible $\sim$10\,m baselines, and in the preceding
discussions we have restricted our attention only to the azimuthally averaged 
visibilities. 
Departures from circular symmetry would be manifest in the visibilities as a modulation
with position angle, while departures from inversion symmetry are indicated by 
non-zero closure phase signals.
As R~Aqr is barely resolved, second-order terms in the visibility function were
hard to extract given the random and systematic errors on the visibility data.
However at all near-IR wavelengths of observation, the visibilities were consistent
with a circular disk, with the upper limits on any ellipticity given by the 
error bars to the circular disk fits given in Figure~1.
With the exception of the 3.08\,$\mu$m observations mentioned below, closure phase
signals were within the error bars of zero, implying no significant departures from
inversion symmetry in the data.

Images of the R~Aqr system in the J band were recently reported by Karovska, McCarthy 
and Christou (1994) from 1991 November speckle observations with the 
MMT (effective aperture 6.86\,m).
These authors found R~Aqr to be dramatically elongated, presenting a size of 60\,mas
along a position angle of 140$^\circ$ while being unresolved in the orthogonal 
direction.  
Unfortunately, there is no evidence for such a signal in our data.
Our 1.236\,$\mu$m data rule out elongations greater than 15\,mas in any direction,
and we note that the expected signal from a 60\,mas elongation would cause our Keck
visibility function to pass beyond the first null -- a radical departure from the
observations.
Unless profound secular changes in the brightness profile occurred between 
1991 and 1999, such elongations appear to be ruled out and we can see of no 
way to reconcile these two datasets.
Full determination of source structure must await the coming generation of 
separated-element imaging interferometers working at substantially higher 
angular resolutions.

It is apparent, from Figure~1, that the 3.08\,$\mu$m data are qualitatively quite
different from the shorter wavelength sets, with significantly larger ($\sim$34\,mas) 
best-fit uniform disk diameters. 
Such dramatic enlargement has been reported before, and seems to be most pronounced
for objects of extremely late spectral type (\cite{Tut_dana99}).
Furthermore, at both 1998 June and 1999 January epochs, the visibilities can be 
seen to be a poor match to the uniform disk profile, implying instead that a 
significant proportion of the flux originates in an extended halo.
Simple geometrical models, consisting of a uniform disk plus an extended Gaussian 
shell, were investigated to see if the 3.08\,$\mu$m visibility functions could be fit.
Good fits to both epochs of data were found for models in which $\sim$20\% of the
flux came from a Gaussian halo of $\sim$100\,mas FWHM. 
Following earlier workers in the optical (\cite{Labeyrie_et77}; \cite{ST87}; 
\cite{HST95}), changes of angular diameter and radial profile
with wavelength can be attributed to the reprocessing of radiation at different
levels in the atmosphere by molecular layers whose opacity characteristics are 
strongly wavelength-dependent.
In contrast to the abundance of prominent molecular features such as TiO and
VO in optical spectra, recent ISO spectra of a number of late M-type stars 
(\cite{Tsuji97}) did not show strong spectral structures near 3.1\,$\mu$m, 
although there was some unaccounted absorption in this region.
Water ice shows an extremely strong absorption at 3.1\,$\mu$m, however this would 
seem to be ruled out on physical grounds and further investigation of the possible 
blanketing effects of common molecules such as CO and H$_2$O is needed.
An alternative scenario is that this extended halo arises from emission from the 
innermost hot circumstellar dust, just beginning to make its presence felt as we 
move towards longer wavelength.

As a final note concerning the 3.08\,$\mu$m data, the closure phase signals
did exhibit a small (few degrees) departure from zero as would be expected if the 
source had a non centro-symmetric brightness distribution.
One way the phase signals could be modeled was to allow the  $\sim$100\,mas 
FWHM Gaussian halo (mentioned above) to move 10\,mas to the southwest with respect 
to the star.
Although this identification of non-centro symmetric elements must be labeled as 
tentative, it is hoped that with the longer baselines available to separated-element
interferometers, any such complexity in this source will be subject to careful scrutiny.

\subsection{Mid-Infrared Visibility Measurements} 

\begin{figure}[ht]
\begin{center}
\epsfig{file=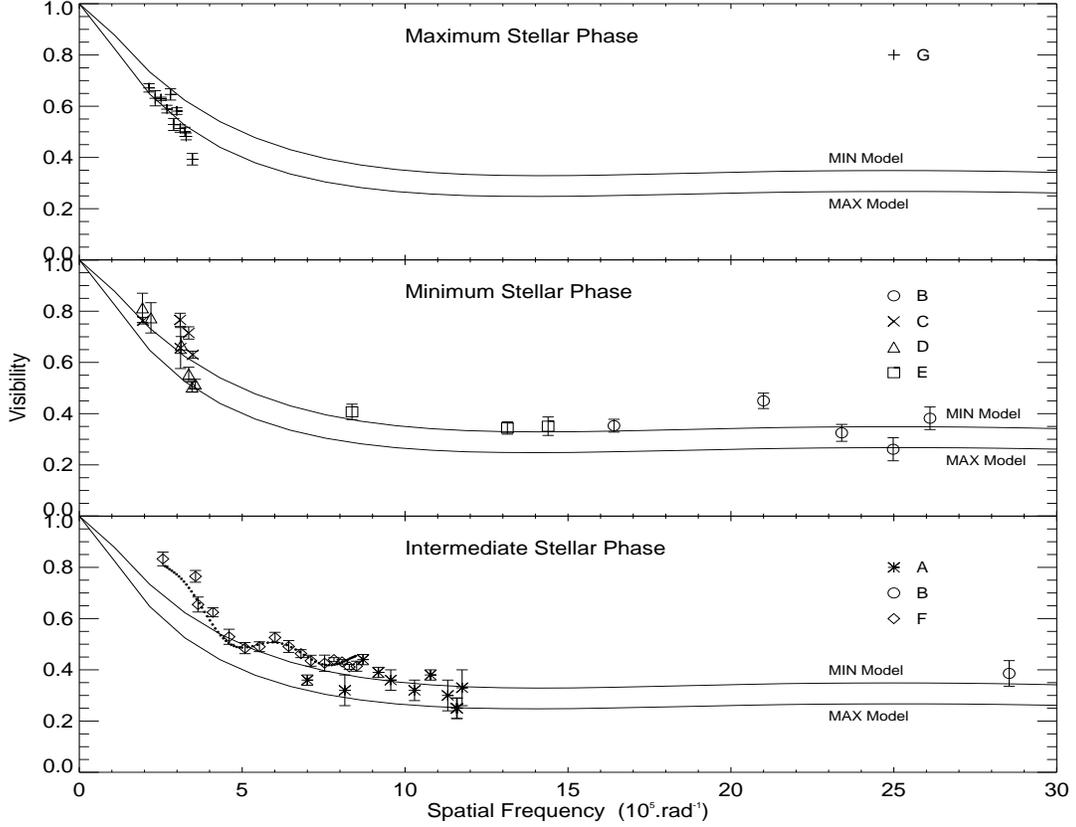,height=11.5cm,width=15cm}
\smallskip
\caption{
Visibility data recorded on R~Aqr from observations at the ISI at 11.15\,$\micron$.
The topmost panel shows only data taken close to stellar maximum 
($\Delta \Phi \le \pm 0.125$), the second panel shows data close to stellar minimum, 
while the third panel shows data taken at intermediate stages in the stellar 
pulsation cycle. 
The various plotting symbols, identified to the right with the letters `A' through
`G', denote different data runs identified in Table~2. 
Overplotted in each panel are two solid lines showing visibility functions from
radiative transfer modeling of the star at stellar maximum and minimum.
Also shown on the third (intermediate) dataset is a dotted line showing a model 
brightness distribution fit to data from set `F' (see text for more details).
\label{fig:midir}}
\end{center}
\end{figure}

Visibilities at 11.15\,$\mu$m, recorded over the seven observing epochs 
(`A' through `G' from Table~2), are plotted in Figure~\ref{fig:midir}.
Although an interferometer baseline is a vector quantity, visibilities are
shown as a function of the (scalar) baseline length. 
This approach is often adopted in cases such as this where coverage of the
UV plane is sparse so as to make full two-dimensional modeling of the visibility
function unwarranted. 

Data are separated into three categories according to the stellar
phase $\Phi$ of the LPV at the time of observation (\cite{aavso99}); 
$\Phi > 0.875$ or $\Phi < 0.125 = $ `Maximum'; $0.375 < \Phi < 0.625 =$ `Minimum'; 
with all other data considered `Intermediate'.
Although the spatial frequency coverage is far from uniform for all stellar 
phases, the general shapes of the visibility
curves are broadly similar at all phases, and they exhibit two clear components.
The relatively rapid drop in visibility at low spatial frequencies argues for
a resolved dust shell component, while the flattening off towards high 
spatial frequencies is indicative of a compact (stellar) component contributing 
some 30$\sim$40\% of the flux.

\begin{figure}[ht]
\begin{center}
\epsfig{file=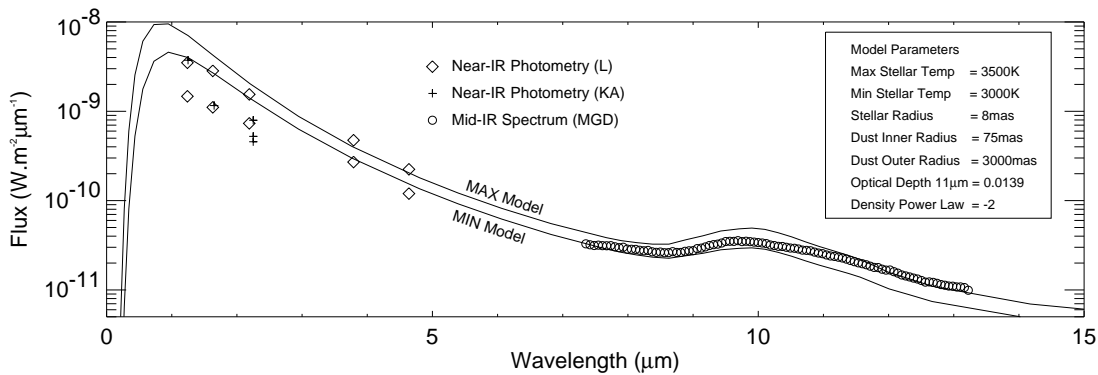,width=15cm}
\smallskip
\caption{
Spectral energy distribution of R~Aqr. The mid-IR spectrum from Monnier et al. 
(MGD 1998) was taken at a stellar phase of 0.46. The near-IR points of Le~Bertre
(L 1993) show points at maximum and minimum light taken from best fits to the
near-IR lightcurve from photometric monitoring data.
The points of Kamath \& Ashok (KA 1999) were all taken fairly close to stellar
maximum (0.85 $\sim$ 0.99).
Overplotted solid lines show numerically computed spectral energy distributions of
radiative transfer models at maximum and minimum light.
The parameters of these models are given in the box to the right.
\label{fig:spect}}
\end{center}
\end{figure}

The physical parameters of the R~Aqr circumstellar environment were modeled
using radiative transfer computations of simulated stellar dust shells 
attempting to match the visibility data, together with published spectrophotometry.
The modeling code used for this purpose, based on the work of 
Wolfire \& Cassinelli (1986), calculates the equilibrium temperature of the 
dust shell as a function of the distance from the star (assumed to be a blackbody).
A wide range of optical properties of dust grains and density distributions of 
dust shells can be modeled, as is described in more detail in Danchi et al. (1994)
and Monnier et al. (1997).
As R~Aqr has an oxygen-rich atmosphere, it is appropriate to use optical 
constants for astronomical silicates while the grain size distribution
follows that of Mathis, Rumpl \& Nordsieck (1977) with dust opacities 
calculated assuming spheroidal Mie scattering using a method
developed by Toon \& Ackerman (1981).
In the absence of detailed multi-wavelength polarization measurements, this
standard model was favored over more complex scenarios such as those with
elongated grains, which should have little effect on the main conclusions
of the paper except perhaps to produce somewhat different optical depths in 
model fits. 
Radiative transfer calculations at 67 wavelengths allow the wavelength-dependent 
visibility curves, the broadband spectral energy distribution, and the 
mid-infrared spectrum all to be computed for comparison with observations.  
Dense sampling of the silicate band in the 10\,$\micron$ region is particularly 
useful for comparison with published spectrophotometry.  
Figure~\ref{fig:spect} gives near- and mid-IR spectrophotometric data taken
from various literature sources for comparison with model spectra.
The optical constants of Draine \& Lee (DL 1984), Ossenkopf, Henning,
\& Mathis (OHM 1992), and David \& Papoular (DP 1990), were all tested in
modeling the dust shell, with the result that OHM (used hereafter) and DP 
constants both provided high-quality fits to the shape of the silicate 
feature, while DL optical constants were not preferred.

For model calculations, the simplifying assumption of uniform isotropic outflow
has been made, resulting in a spherical dust shell with a $\rho \propto r^{-2}$ 
density distribution beginning at a discrete dust condensation radius.
Although there is considerable evidence for departures from spherical symmetry
at larger scales, the primary constraints on the models from the mid-IR 
visibilities and spectra were not extensive enough to warrant more complex
models.

Two models were constructed, one at stellar maximum and one at minimum and
their computed mid-IR visibility functions are given in Figure~\ref{fig:midir}, 
while simulated spectra and physical parameters are in Figure~\ref{fig:spect}.
The diameter of the central star was fixed at 16\,mas (or 3.2\,AU assuming a 
distance of 200\,pc) from near-IR observations
(Section~3.1; \cite{vanB96}) and the temperature fixed to match the mid-IR
luminosity.
As can be seen from Figure~\ref{fig:midir}, the visibility data at maximum and 
minimum light are well fit by the model curves, and we note that all parameters 
of the dust shell model are identical for the two models, with the increase in 
stellar flux being simulated by an elevation in the effective stellar temperature 
of the star by 500\,K.
This in turn leads to an elevation of the temperature of the dust at the inner
radius from 850\,K to 1170\,K.

Serious inadequacies in this simplest-case model scenario are revealed
upon comparison of model predictions with spectro-photometry and 
intermediate-phase visibility data.
As can be seen from Figure~\ref{fig:spect}, the simulated
spectral energy distributions are not well matched to the observations.
In the mid-IR, although the overall flux level and the shape of the model 
silicate feature are in fair accord with the measurements of Monnier et al. 
(1998), the spectral slope towards longer wavelengths is too steep.
This lack of long-wavelength flux argues for the presence of more cool dust,
or a cooler underlying stellar spectrum.
Near-IR photometric measurements (\cite{Lebertre93}; \cite{KA99})
fall well below the models implying increased near-IR opacity, or again a
cooler underlying star; our model overestimates the bolometric luminosity.
However, such decreases in stellar temperature and increases in density of the
circumstellar dust shell were not possible without seriously perturbing the
fits to the visibility data, in particular the relatively high visibilities
recorded at long baselines.

Put simply, the strong point-source (30$\sim$40\%) evident from the mid-IR 
visibility curves drives models to low optical depths and in turn high stellar
luminosities to account for the mid-IR flux (which could otherwise come from
the dust if the optical depth were higher). 
Although a wide range of model parameter space was explored, including variations
on the density-radius law, the inner and outer radii, the dust opacity properties,
and the dust densities, no spherically symmetric model could be found which
overcame the shortcomings with the spectral fitting.
The reader is therefore cautioned against taking the model parameters from
Figure~\ref{fig:spect} as an accurate description of this star; rather they 
describe the best fit to our interferometric dataset which can be provided
by a uniform-outflow spherical dust shell.

There are many imperfections in the modeling process, mostly due to 
oversimplifications or limited knowledge, which could lead to misfitting 
of the data.
Although the use of a blackbody spectrum for the star is common practice, 
evolved stars exhibit significant departures (see for example, \cite{LDB99}).
The optical properties of the dust can also have a significant impact on 
the radiative transfer results, and detailed knowledge of crucial aspects
such as the true grain size distribution are lacking (the dust of \cite{MRN77}
is more appropriate for the ISM).
However, it may be that not all difficulties in fitting may stem from the 
fact that the simple dust radiative transfer models, with available silicate
opacity functions do not give a realistic picture of actual circumstellar
dust shells.

Examination of the intermediate-phase visibilities (bottom panel,
Figure~\ref{fig:midir}) points to an additional possible explanation for the 
poor spectral performance of our models.
While visibilities from data set `A' fall between the Max and Min model
curves as might be expected, those from data set `F' are higher than either,
especially at low spatial frequencies well covered in the other plots.
A likely explanation for this change in visibility can be found by 
examination of Table~2: most of the short baseline data underpinning the
models (sets `C', `D' and `G') along with set `A' were taken at a
position angle on the sky of around $70 \sim 100^\circ$, while the baselines
constituting data set `F' were taken at a very different orientation
($145^\circ$ for the short-baselines).
Hence the difference in these visibilities taken at angles differing by 
$45 \sim 75^\circ$ argues for substantial departures from sphericity.
In such an anisotropic circumstellar environment, the spherical model assumptions
break down and failure to fit to the spectral energy distribution may be the result
of complications such as patchy extinction of the star or radiation leaking 
through holes in the cloud.
Indeed, the extension of asymmetric structure from the well-known nebula down to
successively smaller scales is not surprising, and has already been suggested from
findings of time-variable polarization (\cite{Desh87}).

The visibility data set `F' has rather dense sampling over a range of spatial 
frequencies together with small probable errors, and as is seen from 
Figure~\ref{fig:midir}, it exhibits wiggles or bumps (e.g. at 
$6 \times 10^5$\,rad$^{-1}$) which indicate departure from a simple 
smooth radial intensity distribution. 
In order to assess this further evidence for anisotropy in the R~Aqr dust shell,
model brightness distributions (not, in this case, based on radiative transfer 
computations) have been fit, the result of which is overplotted as a dotted line 
in panel~3 of Figure~\ref{fig:midir}.
The stellar component of our best-fit model contributed 36\% of the flux, with 61\% 
coming from a spherical circumstellar dust shell assumed to have a simple Gaussian 
profile with a FWHM of 350\,mas. 
A third component, modeled as a localized point-source contributing 3\% of the 
total flux and located 700\,mas from the star at a position angle of $100^\circ$ 
(or $-80^\circ$), produces the wiggles noted in the data.
Although such a feature might be some localized concentration of dust, or possibly
a local warming, serious discussion should be deferred until 
more complete Fourier coverage can be obtained.
With the sparse coverage afforded by the single interferometer baseline, 
a host of different models could be devised in which some additional flux is 
originating at some distance from the star and thereby generating the high-frequency
signal we see. 
Therefore the likely significance to be placed upon our model parameters is low.
This discussion is included simply as additional evidence for departure from a uniform
outflow, and as an encouragement to further efforts at full high angular resolution 
imaging of this system in the mid-IR.

When non-spherical dust halos occur, it is rarely possible to discriminate 
between a number of viable outflow models when only relatively limited Fourier
data are available, even though reasonable models may be chosen 
(e.g. \cite{Lopez_et97}). 
The simple isotropic outflow model given here provides a reasonable fit to most 
of our mid-IR visibility data.
However, deviations from the model are clear, and the more involved task of mapping 
and fitting of spectral data await the recovery of the full two-dimensional 
complex visibility function.
For this reason, a detailed comparison with other models such as the two-shell 
model of Anandarao \& Pottasch (1986) or the uniform outflow model of 
Le Sidaner \& Le Bertre (1996) is not given here. 
However, our inner dust shell radius is in agreement with results of these 
earlier workers (for \cite{AP86}, we consider only their hot inner shell which should 
dominate over the cool outer shell ($T_2 = 87.1$\,K) in this wavelength region).
On the other hand, significant differences include lower optical depths and hotter 
stellar temperatures. 

All spherically symmetric models investigated have been found to be inadequate in 
fitting the expanded dataset encompassing near- and mid-IR interferometry and the
spectral energy distribution from the near- to the far-IR.
This finding is in agreement with studies of $o$~Ceti, another symbiotic system 
observed interferometrically in the mid-IR \cite{Lopez_et97}. 
With further work at high resolutions, particularly directed towards full imaging
of these systems, the effects of the companion's presence, both gravitational and 
radiative, in the shaping of the circumstellar dust shell may be elucidated.

%___________________________________CONCLUSIONS________________________
\section{Conclusions}

High-resolution interferometric studies of R~Aqr at narrow bandwidths
within the J, H and K bands find no evidence for any significant departure from 
the best-fitting model of a marginally-resolved stellar disk, which we identify
with the LPV component of the system.
Any companion present and separated by more than $\sim 15$\,mas must exhibit 
a magnitude difference in excess of  $\Delta M $\simge~5\,mag in the near infrared.
These results are consistent with those of Van Belle et al. (1996), but
not with Karovska et al. (1994).
An enlargement, by approximately a factor of 2, is reported for the
apparent size at a wavelength of 3.08\,$\micron$ as compared with shorter
near-infrared bands, which is attributed to molecular blanketing by an 
unidentified species in the atmosphere, or thermal emission from material 
in an extended halo.
Tentative evidence for asymmetry at this wavelength is also reported.
In the mid-IR, visibility data obtained with the ISI have been used to
constrain simple uniform spherical outflow models.
Using self-consistent radiative transfer calculations, good fits to most
of the visibility data have been obtained.
However, serious shortcomings were found when comparing the synthetic spectral
energy distributions to measurements, many of which may be attributed to 
inadequate knowledge of the detailed dust shell parameters.
Furthermore, mid-IR interferometry obtained at a different position angle
on the sky points towards substantial departures both from spherical symmetry and
from a simple radial distribution. 
These indicate that the shortcomings of the present models are probably in large
part be due to the oversimplistic assumption of an isotropic uniform outflow.
Further high-resolution studies, both in near-infrared with higher spatial
resolution and dynamic range, and in the mid-infrared with more complete 
Fourier coverage, are important in order to fill in the gaps in 
understanding of this star.

%___________________________________ACKNOWLEDGEMENTS___________________
\acknowledgments
{ 
We thank Everett Lipman for help in securing many of the ISI measurements 
published herein, and for developmental work on the instrument.
Long-baseline interferometry in the mid-infrared at U.C. Berkeley is 
supported by the National Science Foundation 
(Grants AST-9315485, AST-9321289, AST-9500525, \& AST-9731625) and by the 
Office of Naval Research (OCNR N00014-89-J-1583  \& FDN0014-96-1-0737). 
Some of the measurements herein were obtained at the W.M. Keck Observatory, 
made possible by the generous support of the W.M. Keck Foundation, 
and operated as a scientific partnership among the California Institute 
of Technology, the University of California, and NASA.

} 
% \pagebreak

%___________________________________BIBLIOGRAPHY_______________________

%\end{references}

\end{document}